\PassOptionsToPackage{table,xcdraw,dvipsnames}{xcolor}
\documentclass[sigconf]{acmart}
\usepackage{booktabs}
\usepackage{caption}
\usepackage{subcaption}
\usepackage{booktabs}
\usepackage{arydshln}
\usepackage{multirow}
\usepackage{amsmath}
\usepackage{float}
\usepackage{makecell}
\usepackage{pythonhighlight}

\newcommand{\set}[1]{\mathcal{#1}}

\newcommand{\tracks}{\texttt{Tracks}}
\newcommand{\podcasts}{\texttt{Podcasts}}
\newcommand{\books}{\texttt{Books}}

\newcommand{\clicks}{\texttt{Click}}
\newcommand{\qgen}{\texttt{QGen}}
\newcommand{\inpars}{\texttt{InPars}}
\newcommand{\cqg}{\texttt{CtrlQGen}}

\newcommand{\broad}{\texttt{broad}}
\newcommand{\narrow}{\texttt{narrow}}

\newcommand{\podcastbroad}{\texttt{Podcasts$_{\broad{}}$}}
\newcommand{\tracksbroad}{\texttt{Tracks$_{\broad{}}$}}

\newcommand{\narrowcols}{\textit{narrow-fields}}
\newcommand{\broadcols}{\textit{broad-fields}}
\newcommand{\broadcolsft}{\textit{broad-fields-ft}}

\newcommand{\bm}[1]{\texttt{BM25}}
\newcommand{\biencoder}[1]{\texttt{Bi-Encoder}}

\newcommand{\weaklabelsun}[1]{\texttt{WeakLabeling-Un}}
\newcommand{\weaklabelsip}[1]{\texttt{WeakLabeling-IP}}

\AtBeginDocument{%
  \providecommand\BibTeX{{%
    \normalfont B\kern-0.5em{\scshape i\kern-0.25em b}\kern-0.8em\TeX}}}

\copyrightyear{2023}
\acmYear{2023}
\setcopyright{acmlicensed}\acmConference[WWW '23]{Proceedings of the ACM Web Conference 2023}{May 1--5, 2023}{Austin, TX, USA}
\acmBooktitle{Proceedings of the ACM Web Conference 2023 (WWW '23), May 1--5, 2023, Austin, TX, USA}
\acmPrice{15.00}
\acmDOI{10.1145/3543507.3583261}
\acmISBN{978-1-4503-9416-1/23/04}

\acmPrice{15.00}
\acmISBN{978-1-4503-XXXX-X/18/06}

\begin{document}

\title{Improving Content Retrievability in Search with Controllable Query Generation}

\author{Gustavo Penha$^1$, Enrico Palumbo$^2$, Maryam Aziz$^3$, Alice Wang$^3$, Hugues Bouchard$^4$}
\affiliation{
\institution{Spotify}
\country{$^1$Netherlands, $^2$Italy, $^3$USA, $^4$Spain}
}
\email{{gustavop,enricop,maryama,alicew,hb}@spotify.com}  






\renewcommand{\shortauthors}{}

\begin{abstract}


An important goal of online platforms is to enable content discovery, i.e. allow users to find a catalog entity they were not familiar with. A pre-requisite to discover an entity, e.g. a book, with a search engine is that the entity is \emph{retrievable}, i.e. there are queries for which the system will surface such entity in the top results. However, machine-learned search engines have a high retrievability bias, where the majority of the queries return the same entities. This happens partly due to the predominance of narrow intent queries, where users create queries using the title of an already known entity, e.g. in book search ``\textit{harry potter}''. The amount of broad queries where users want to discover new entities, e.g. in music search ``\textit{chill lyrical electronica with an atmospheric feeling to it}'', and have a higher tolerance to what they might find, is small in comparison.  We focus here on two factors that have a negative impact on the retrievability of the entities (I) the training data used for dense retrieval models and (II) the distribution of narrow and broad intent queries issued in the system. We propose \cqg{}, a method that generates queries for a chosen underlying intent---narrow or broad. We can use \cqg{} to improve factor (I) by generating training data for dense retrieval models comprised of diverse synthetic queries. \cqg{} can also be used to deal with factor (II) by suggesting queries with broader intents to users. Our results on datasets from the domains of music, podcasts, and books reveal that we can significantly decrease the retrievability bias of a dense retrieval model when using \cqg{}. First, by using the generated queries as training data for dense models we make 9\% of the entities retrievable---go from zero to non-zero retrievability. Second, by suggesting broader queries to users, we can make 12\% of the entities retrievable in the best case.
\end{abstract}

\maketitle

\section{Introduction}

\definecolor{Mycolor}{HTML}{a3a3a3}
\definecolor{Mycolor2}{HTML}{fc0adc}

\begin{figure}[]
    \centering
    \includegraphics[width=0.42\textwidth]{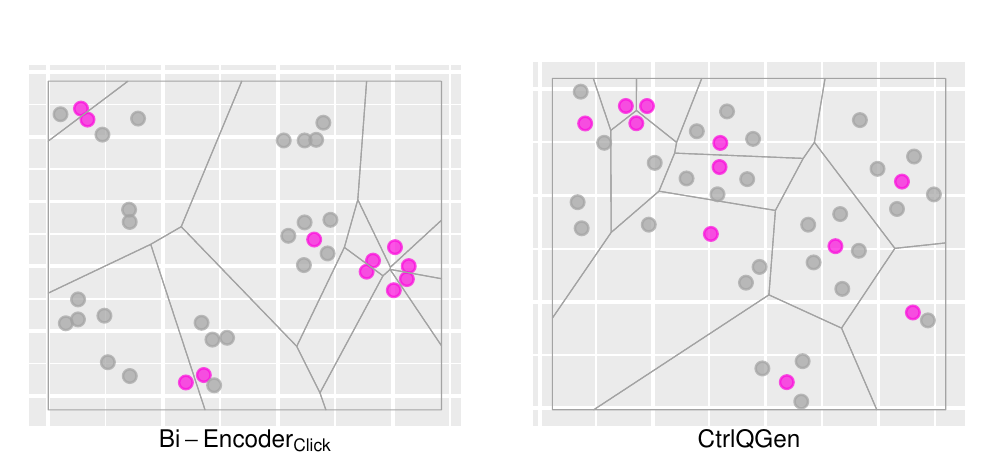}
    \caption{TSNE reduction of \textcolor{Mycolor}{queries} and \textcolor{Mycolor2}{entities} when embedded with a Bi-Encoder trained with query logs and clicked entities (left), and synthetic queries from our proposed method \cqg{}  (right). The left model surfaces the same four entities for most queries while six entities are never retrieved as the most similar entity. The right model distributes the queries better, i.e. has less retrievability bias.}
    \label{fig:intro_image} 
\end{figure}
 

In online content platforms, users can search for catalog entities\footnote{Catalog entities are items from a platform that can be retrieved and/or recommended to users. For example, the book \textit{``The Fellowship of the Ring by J.R.R Tolkien''} is an entity from an online book platform. We refer to such items as \emph{entities} throughout the paper.} that they are already familiar with, for example, they issue queries with the title of a track to listen to next or of a book they would like to read. This type of search based on bibliographic data (e.g. title, artist, author, etc) to find entities~\cite{bainbridge2003people} has been referred to as \emph{narrow} intent queries~\cite{hosey2019just}. However, user information needs are diverse and can be more complex depending on their current mindset~\cite{li2019search}. 

When users have an exploratory mindset, they have a higher tolerance and are prone to explore different alternatives through \emph{broad} queries. Non-focused information needs are generally complex and require multiple interactions. Many users solve such information needs outside the search engine of the platforms, by asking broad queries to other users in forums such as subreddits\footnote{See for example \url{https://www.reddit.com/r/musicsuggestions/} or /r/booksuggestions/.} as existing search systems are ineffective for broader intents. 

Broad intents are an opportunity to surface under-served entities that would not be discovered otherwise without affecting user satisfaction~\cite{tomasi2020query}. While approaches to promote the discovery of entities have been studied from the perspective of recommender systems~\cite{aziz2021leveraging,mehrotra2021algorithmic}, they do not generalize to search engines where there is an input query. A prerequisite to improving the discoverability of entities through search is that the entity is retrievable. ~\citet{azzopardi2008retrievability} defined the \emph{retrievability of a document as how many queries lead to the entity being surfaced in the top-k results.} 


For example, if we assume that the users will only interact with the top-1 ranked entity of the list, the dense retrieval model used to embed queries and entities in the left Voronoi plot of Figure~\ref{fig:intro_image} would make one of the five leftmost entities appear for every query (they are closest neighbors in the embedded space). These entities would have a high concentration of retrievability, i.e. retrievability bias, compared to the remaining entities which have no query close in the embedding space. Retrievability bias limits exploration, as it becomes harder to discover new entities through search when they have low retrievability scores.

In this paper, we study the effect of \emph{generating queries} on the retrievability of the system. Although the implications of query generation techniques for training document/passage dense models have been studied in detail~\cite{liang2020embedding,ma2020zero,wang2021gpl,dai2022promptagator}, little attention has been given to generating queries for \emph{entities} and their impact on the effectiveness and \emph{retrievability bias}. Dense models have shown promising results for different retrieval tasks~\cite{lin2021pretrained}, requiring a significant amount of in-domain supervision data for training~\cite{thakur2021beir}. Query generation approaches have shown to be effective in generating training data for domains with a scarcity of labeled data~\cite{liang2020embedding,ma2020zero,bonifacio2022inpars}.



Unlike previous approaches to query generation which are agnostic to search intents, we propose \cqg{} which controls for the underlying intent. By generating both narrow and broad queries for an entity we are able to (I) train the dense retrieval model for both types of intents and (II) suggest broader and more exploratory queries to users. With the use of weak supervision through the proposed \emph{weak labeling functions}, \cqg{} does not strictly require any training data to generate synthetic queries for a given entity.
 
With our empirical evaluation using three datasets in the domains of music, podcasts, and books we set out to answer the following research question: \emph{\textbf{To what extent can we reduce the retrievability bias of entity search with automatically generated queries without significant impact in the effectiveness?}} 


Considering that the retrievability of an entity depends on (I) the retrieval model which decides which entities are surfaced for each query and (II) the set of queries used for the estimation, we generate two retrievability debiasing hypotheses that focus on modifications to the retrieval model and the set of queries respectively. Our first hypothesis, \textbf{H1}, is that training dense retrieval models with \cqg{} queries will lead to less retrievability bias compared to training with real queries and their respective clicked entities. The click data is prone to different biases, for example, many queries will be issued for the most popular entities, i.e. popularity bias, and after training the model on such data and this bias will be reinforced in later interactions with the system. Conversely, with \cqg{} we can obtain pairs of query-entity to train the model for any given entity, which can be randomly sampled from the collection. 

Our second hypothesis, \textbf{H2}, is that suggesting broad queries using \cqg{} will lead to less retrievability bias. Narrow queries have by definition less relevant entities than broad queries. By assisting users in formulating their queries with the suggestion of broad queries we can potentially influence users' query behaviors and then have an impact on the query type distribution.

Our main findings and contributions are:
\vspace{-0.1cm}
\begin{itemize}
    \item We introduce \cqg{}, a novel method to generate queries for a given entity conditioned on a desired underlying intent (\narrow{} or \broad{}). We demonstrate two ways of using the generated queries: as training data for dense retrieval models and as query suggestions.

    \item We find positive evidence for \textbf{H1}: dense models fine-tuned on synthetic queries have significantly less retrievability bias than models fine-tuned on click data. When using the queries from the proposed \cqg{} we reduce the retrievability bias by 10\% in terms of Gini scores on average when compared to a model that uses the click data and make 9\% of the \tracks{} collection of entities retrievable---go from zero to non-zero retrievability score. 
    
    \item Regarding \textbf{H2}, we show that applying \cqg{} for generating query suggestions can reduce the retrievability bias of the system up to 9\% percent and increase the number of entities that have non-zero retrievability 11\% for the \tracks{} collection when using a Bi-Encoder model that was trained with an unbiased set of queries.
\end{itemize}

Next, we describe the related work, followed by the proposed method in Section 3. Section 4 describes the experimental setup used to answer the research question, followed by our experiments in Section 5. We conclude the paper in Section 6.


\section{Related Work}
We first discuss here related work on search for the domains considered here. Then we look into retrievability followed by a discussion on query generation techniques and their applications.

\subsection{Entity Search}

A number of studies have explored user behavior when searching for specific entities such as music tracks, products, and books~\cite{hosey2019just,garcia2018understanding,bainbridge2003people,laplante2008everyday}. While focused searches have the goal of finding a specific entity, non-focused searches involve broader intents, where the user is in an exploratory mindset~\cite{li2019search,su2018user,sguerra2022navigational}. 

The Social Book Search Lab CLEF~\cite{koolen2016overview} that ran from 2011 to 2016\footnote{The book corpus of the SBS tasks is no longer available.} enabled a number of studies in complex search for the book domain ~\cite{ullah2020social,bogers2017supporting,bogers2018m,ullah2021improving,chaa2018combining}. A richer document representation for books which contains for example reviews, tags, and controlled vocabulary was shown to have better retrieval effectiveness.  It has also been shown in the music domain that multiple sources of data such as metadata, audio features, tags, and lyrics lead to better effectiveness for downstream tasks~\cite{kim2020one}. For podcast search, the TREC2020 podcasts track~\cite{jones2021trec} revealed that adding the additional information of transcripts also leads to higher effectiveness when compared to using only the episode title and description. 

Another external source of information for entities that was shown to be useful for downstream tasks~\cite{stamatelatos2021point} is the concept of lists, where users group together a number of entities that are similar in a way. In the music domain, this is often referred to as playlists. The creation of lists with curated entities is also common in the domains of books~\cite{liu2014recommending} and movies~\cite{greene2013discovering}.


\begin{figure*}[ht!]
    \centering
    \includegraphics[width=1\textwidth]{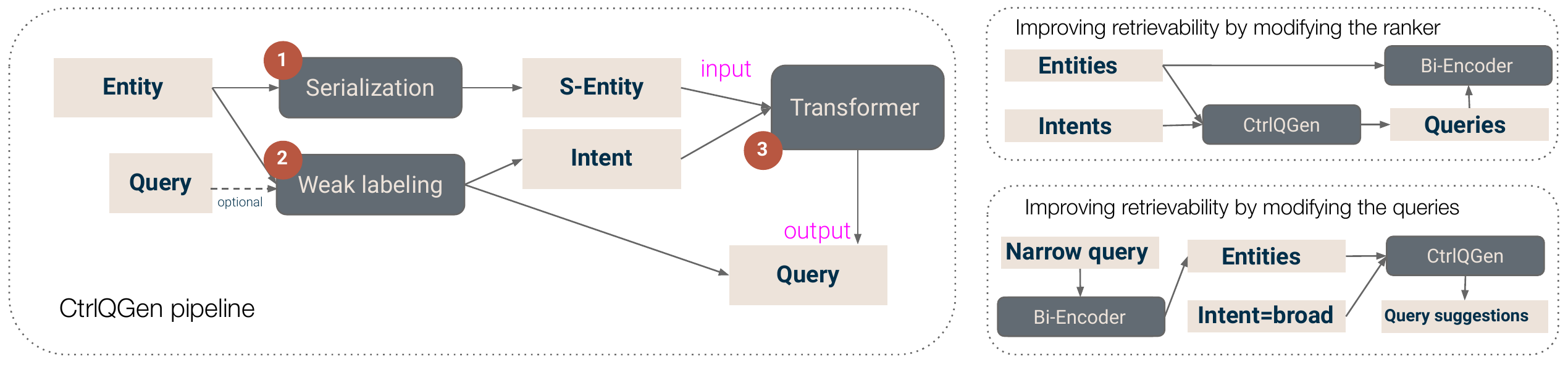}
    \caption{\textcolor{Brown}{Left}: Components of the \cqg{} method. Each entity is serialized by concatenating the values of each metadata, e.g. \textit{title: The Fellowship of the Ring [SEP] author names: J.R.R. Tolkien [...]}. (2) Labeled data (\textit{entity ; query ; intent}) is not strictly required, due to the use of weak labeling functions which output a query and intent for a given entity, e.g. (\textit{The Fellowship of the Ring ; fantasy book ; \broad{}}). (3) Control over the underlying intent (\narrow{} or \broad{}) when generating the query via prompting, e.g. ``\textit{Generate a query with \narrow{}/\broad{} intent from: <serialized\_entity>}''. \textcolor{Brown}{Right}: Different ways of using the proposed \cqg{} to improve the retrievability of the search system: modifying the ranker by fine-tuning on synthetic queries and modifying the set of queries by suggesting broad queries for the narrow intent queries issued.}
    \label{fig:model_diagram} 
\end{figure*}

\subsection{Retrievability}


To estimate the retrievability of a document~\cite{azzopardi2008accessibility,azzopardi2008retrievability} proposed to sum the popularity of the queries that retrieve the given document above a position that the user would actually look at (e.g. in the top-5 documents). Retrievability scores can be used to determine if a retrieval system has a concentration of retrievability, for example, to verify if certain types of documents are being surfaced more than others. For example,~\cite{roy2022studying} showed that for a collection with datasets and articles, the retrievability bias was stronger for datasets when compared with articles. 

Even though a system with less retrievability bias does not necessarily mean that the system is more effective, studies have found a correlation between the two~\cite{bashir2017retrieval,wilkie2013relating,wilkie2014retrievability}, suggesting that a measure of retrievability bias can potentially be used to select better retrieval systems. In order to reduce the retrievability bias of a system ~\cite{bashir2010improving} proposed a query expansion technique with a novel document selection process for pseudo-relevance feedback in the domain of patent search.~\citet{chakraborty2020retrievability} proposed to use retrievability of a document over a set of query variations to decide which documents to use for relevance feedback. Finding which queries lead to a document can be also used to improve search transparency~\cite{li2022exposing}.

\subsection{Query Generation}
Query generation techniques can be broadly categorized based on their input: documents or queries. For generating known-item \textbf{queries for a given document}, i.e. queries where the task is to find a previously seen document, techniques have been proposed that select a number of document terms based on different sampling methods~\cite{kumar2011algorithmic,azzopardi2007building,azzopardi2006automatic}. ~\citet{liu2021strong} tackled a similar problem with the additional constraint that the generated queries are also informative. Generating queries for a given document using a seq2seq model was first proposed by~\cite{nogueira2019document}. Unlike sampling methods proposed for generating known-item queries, a seq2seq approach such as \textit{docT5query}~\cite{nogueira2019doc2query} is able to generate queries where its terms do not occur in the input document, being able to mitigate the vocabulary mismatch problem. Similarly, ~\cite{liang2020embedding,ma2020zero,wang2021gpl} generate queries based on documents using a transformer encoder-decoder model, but instead of using the queries for document augmentation, they employ the queries as additional training data for training bi-encoders, leading to significant gains in retrieval effectiveness---specially in cross-domain evaluation settings. It has also been proposed to replace the fine-tuned encoder-decoder model to generate queries with little supervision by doing in-context learning with models such as GPT-3~\cite{dai2022promptagator,bonifacio2022inpars}.~\citet{zhuang2022bridging} used generated queries with the goal of improving the effectiveness of the emerging differentiable search indexes. Another recent direction for query generation is to incorporate explicit knowledge when generating queries, e.g. with the use of knowledge graphs~\cite{shen2022diversified,cho2022query,han2019inferring}.


Generating \textbf{queries for a given query} has also been shown to be useful in IR. For example by generating query suggestions or reformulations that help users explore and express their information needs~\cite{cao2008context,mei2008query}. Another objective is to generate query variants that can be used to obtain more effective ranking models by combining such variants for the given query~\cite{belkin1995combining,benham2019boosting}, and also to better evaluate ranking models~\cite{penha2022evaluating,zuccon2016query,bailey2017retrieval}.

The closest to our problem is the generation of queries for product search.~\citet{lien2022leveraging} used textual data from the reviews associated with the documents (products) to generate queries automatically for the following products: headphones, tents, and conditioners. In the domain of movies,~\citet{10.1145/3404835.3463071} generated queries automatically for a document (a movie) based on a number of predefined semantic components such as genre and year. We propose here a method to generate queries that can take advantage of manually created functions as a weak supervision signal, and also employ pre-trained language models. Unlike previous methods to generate queries, \cqg{} does intent-aware generation, where it is possible to control for the underlying intent of the output query.
\section{Controllable Query Generation}

In this section, we first describe the three main components of the proposed \cqg{} followed by different applications for the generated queries. Figure~\ref{fig:model_diagram} displays a diagram of the model, as well as two different ways to employ the synthetic queries. The serialization module is required to obtain a text representation for a given entity so that text-based models can use that as input. The second component is called weak labeling, which is able to bypass the need for a large amount of labeled data. Finally, the last component is the intent-aware generation, which is able to control for different types of intent (broad and narrow).

\subsection{Model Components}

\subsubsection{Serialization}
This module takes as input an entity $e$ and outputs a string representation of the entity: $e_{serialized} = s(e)$. The serialization function $s$ concatenates every metadata column of the entity with their respective values, using a special token: $ s(e) = \textcolor{brown}{col_{1}}: val_{1} [SEP] \textcolor{brown}{col_{2}}: val_{2} [SEP] ... [SEP] \textcolor{brown}{col_{n}}: val_{n}$. So for example the book with the title \textit{The Fellowship of the Ring} becomes:

    \vspace{4mm} 
     \fbox{\begin{minipage}{23em}
    \textit{\textcolor{brown}{title}:  The Fellowship of the Ring [SEP] \textcolor{brown}{series name}: The Lord of the Rings \#1 [SEP] \textcolor{brown}{author names}: J.R.R. Tolkien [SEP] \textcolor{brown}{publication year}: 1954 [SEP] \textcolor{brown}{language}: EN [SEP] \textcolor{brown}{genres}: Fantasy, Classics, Fiction, Adventure, High Fantasy, ...[SEP] \textcolor{brown}{description}: One Ring to rule them all, One Ring to find them ...[SEP] \textcolor{brown}{review}: This book is full of wonder and adventure with fantastic writing ... [SEP] \textcolor{brown}{lists}: fantasy, uk-and-ireland, witches-wizards, fiction, british, ...}
    \end{minipage}}
    \vspace{4mm}

\subsubsection{Weak Labeling}
In order to train \cqg{} we require a dataset $\set{D}=\{(e_i, i_i, q_i)\}_{i=1}^{M}$ with training triplets of entity, intent, and query, which are the input, control variable, and output respectively. In each triplet, the query $q$ has the underlying intent $i$ (\narrow{} or \broad{}) when matching with the entity $e$. One option to acquire such data is to ask annotators to create \narrow{} and \broad{} queries for a given entity. Alternatively, we can employ weak labeling functions that generate such data based on heuristics.

We present here two flavors of weak labeling functions. The first is completely \emph{unsupervised} (\weaklabelsun{}), and thus is able to generate both query and intents for any given entity. The second requires queries that are related to each entity and thus is based on \emph{intent prediction} of the given query (\weaklabelsip{}).

\begin{figure}[H]
    \centering
    \includegraphics[width=0.50\textwidth]{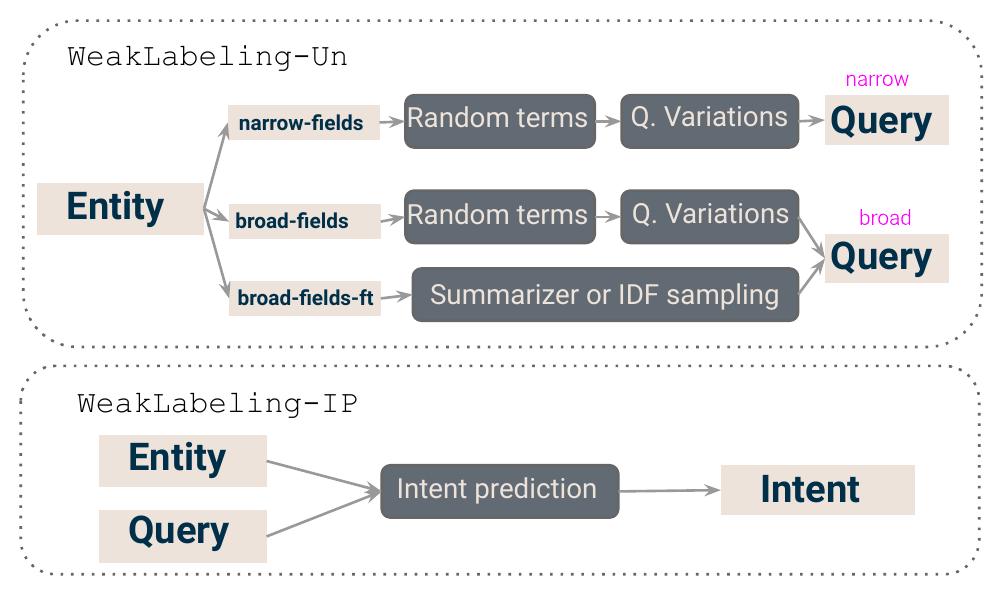}
    \caption{Two variants of the weak labeling functions. While \weaklabelsun{} (top) outputs query and intents for a given entity, \weaklabelsip{} (bottom) outputs an intent label for a given pair of entity and query. }
    \label{fig:weak_labeling_diagram} 
\end{figure}

\textbf{\weaklabelsun{}}. The core intuition is that we can define a set of metatada columns that are inherently associated with narrow intent queries since they can identify the entity (\narrowcols{}), e.g. title and artists, and a set of metadata columns that capture characteristics of the entity that other entities might also have, e.g. genres, and thus can be considered to be broad columns (\broadcols{}). In order to generate a set of queries and intents for a given entity we rely on randomly sampling terms from all possible combinations of the respective fields. So for example to generate a narrow intent query in the music domain, we could use either the title of a track, the album, the artist, or a combination of the three. After sampling terms from such the respective columns for the query, we apply a number of functions to generate query variations in a stochastic manner: shuffling words, adding misspellings, and removing prefixes\footnote{Since the datasets considered come from a large-scale online platform with \emph{instant search}, many log queries are not complete and are just prefixes of the entity titles. This happens because the user might stop before the end of the query as the result could be already found in the list of results.}. 

Specifically, when generating broad queries, we differentiate between metadata columns that are based on free text (\broadcolsft{}), e.g. reviews, and the ones which are already category-like terms (\broadcols{}), e.g. genres. For the free text columns, in order to avoid selecting terms that are uninformative, we apply a sampling strategy that prioritizes terms with higher IDF. As another weak labeling function for the free text columns, we apply a text summarization model to select more informative terms.

\textbf{\weaklabelsip{}} In order to take advantage of existing data of entities and queries, e.g. query logs with clicked entities, this variant predicts if the query is \broad{} or \narrow{} based on its narrow and broad columns. If the similarity\footnote{We employ here a transformer sentence representation and cosine similarity.} of the query and the values of the narrow queries is higher than the similarity of the query with the values of the broad queries then the weak label will be \narrow{}, otherwise \broad{}. So for example, if the entity is a book with the title ``\textit{The Brothers Karamazov}'', and the input query is ``\textit{Karamazov}'', the label would be \narrow{} whereas if the input query is ``\textit{russian theological fiction}'' the label would be \broad{} as it would be more similar to the categories of the book. 

\subsubsection{Intent-aware Generation}
Given the training dataset $\set{D}=\{(e_i, i_i, q_i)\}_{i=1}^{M}$, we train an encoder-decoder model $G$ that receives as input the entity and the underlying intent to control for, and it outputs the query: $G(e,i) = q$. In order to achieve that we rely on adding the control variable as part of the language model prompt. We train the model with the following prompt: ``\textit{Generate a query with \narrow{}/\broad{} intent from: <serialized\_entity>}'' and its respective query as the output. So for example the query ``\textit{lord of th}'' with intent \narrow{} would lead to the following training instance:
    
    \vspace{4mm} 
     \fbox{\begin{minipage}{23em}
    \textit{\textbf{Input} Generate a \textbf{narrow} query from: \textit{\textcolor{brown}{title}:  The Fellowship of the Ring [SEP] \textcolor{brown}{series name}: The Lord of the Rings \#1 [SEP] \textcolor{brown}{author names}: J.R.R. Tolkien [SEP] \textcolor{brown}{publication year}: 1954 [SEP] \textcolor{brown}{language}: EN [SEP] \textcolor{brown}{genres}: Fantasy, Classics, Fiction, Adventure, High Fantasy, ...[SEP] \textcolor{brown}{description}: One Ring to rule them all, One Ring to find them ...[SEP] \textcolor{brown}{review}: This book is full of wonder and adventure with fantastic writing ... [SEP] \textcolor{brown}{lists}: fantasy, uk-and-ireland, witches-wizards, fiction, brittish, ...}}
    
    \textbf{Output} \textit{lord of th}
    \end{minipage}}
    \vspace{4mm}

\subsection{Applications}

\subsubsection{Synthetic training data}
We can use the generated queries to train Bi-Encoder retrieval models, as shown on the right top part of Figure~\ref{fig:model_diagram}. For a randomly sampled set of entities $\set{E'}$ from the collection $\set{E}$ we apply \cqg{} with both desired intents $q'_{narrow} = G(e, \narrow{})$ and $q'_{broad} = G(e, \broad{})$ for each $e$ in $\set{E'}$. After that, given a desired weight proportion of broad queries and narrow queries ($P_{narrow}$, $P_{broad}$) we can sample training instances from the synthetic generated queries $\set{Q'}$ for training the Bi-Encoder. This gives us a dataset of pairs of synthetic queries and respective relevant entities that can be used to train Bi-Encoder models, controlling for the desired proportion of underlying intents.

\subsubsection{Query suggestion} We can employ the \cqg{} model to perform query suggestion, as shown on the bottom right part of Figure ~\ref{fig:model_diagram}. Since the majority of the queries for entities have a narrow intent behind them, one approach to modifying the user's behavior is to suggest broader queries. In order to do that, we can employ the generated queries in the following manner. First, for a given input query $q$, we can obtain a list $\set{R}_q$ with the top-k entities ranked for it using a ranking model. For each entity in the top-k ranked list, we apply \cqg{} to generate a set of broad queries $\set{Q'}$ to recommend: $\set{Q'} = \{G(e_{i}, \broad{})\}$ for $e_{i}$ in $\set{R}_q$\footnote{In our experiments this set of queries $\set{Q'}$ is appended, according to a percentage of acceptance, to the set of log queries $\set{Q}$ in order to calculate the retrievability bias.}. The complexity of this approach is $O(n^2 *d * k)$, where $n$ is the sequence
length, $d$ is the number of dimensions of the transformer model and $k$ is the size of the list considered to generate suggestions.

\begin{table}[ht!]
\footnotesize
\caption{Datasets metadata and statistics. Metadata columns 1--3 are considered to be \narrowcols{}, whereas 4--9 are \broadcols{}. In the experiments the broad columns which are free-text (\broadcolsft{}) are: Episode \& show description and Transcript for \podcasts{}, and User reviews and Description for \books{}.}
\label{table:dataset_statistics}
\begin{tabular}{@{}lllll@{}}
\toprule
 &  & \tracks & \podcasts & \books \\ \midrule
Metadata & \begin{tabular}[c]{@{}l@{}}(1)\\ (2)\\ (3)\\ (4)\\ (5)\\ (6)\\ (7)\\ (8)\\ (9)\end{tabular} & \begin{tabular}[c]{@{}l@{}}Title\\ Album name\\ Artist names\\ Release year\\ Language\\ Genres\\ Descriptors\\ Lyric\\ User Playlists\end{tabular} & \begin{tabular}[c]{@{}l@{}}Title\\ Show name\\ Host names\\ Ingested date\\ Language\\ Categories\\ Episode \& show description\\ Transcript\\  Topics\end{tabular} & \begin{tabular}[c]{@{}l@{}}Title\\ Series name\\ Author names\\ Publication year\\ Language\\ Genres\\ Description\\ User reviews\\ User lists\end{tabular} \\ \midrule
\multicolumn{2}{l}{\# docs} & 682k & 600k & 617k \\ \midrule
\multicolumn{2}{l}{\# queries} & 100k & 100k & 100k \\ \midrule
\multicolumn{2}{l}{\begin{tabular}[c]{@{}l@{}}\clicks{} \# qrels  \\ train/val/test\end{tabular}} & 75.9k/9.5k/9.5k & 14.4k/1.8k/1.8k & 117.5k/14.7k/14.7k \\ \midrule
\multicolumn{2}{l}{Avg doc len} & 55.87 & 80.76 & 161.58 \\ \midrule
\multicolumn{2}{l}{Avg query len} & 1.96 & 3.06 & 4.47 \\ \bottomrule
\end{tabular}
\end{table}
\section{Experimental Setup}

In this section, we first describe the data used to test our hypotheses, followed by the implementation details of the methods and baselines as well as how we evaluate different approaches.

\subsection{Datasets}
In order to test our hypothesis and compare different methods to generate queries we rely on three datasets: \tracks{}, \podcasts{}, and \books{}. For each dataset, we have a set of entities (>600k entities), a set of 100k queries, and a set of relevance judgements. Table~\ref{table:dataset_statistics} describes the statistics of the datasets and examples of entities.

While the queries and entities from \textbf{\tracks{}} and \textbf{\podcasts{}} were extracted from a large-scale online platform the \textbf{\books{}} dataset is a subset of the Goodreads public dataset from~\cite{DBLP:conf/recsys/WanM18}~\footnote{\href{https://github.com/MengtingWan/goodreads}{https://github.com/MengtingWan/goodreads}}. The query sets from \tracks{} and \podcasts{} are a unique subset of randomly sampled entities and queries from the logs of a large scale online audio platform, where clicks for a given entity after issuing the query are considered to be the relevance signal in our experiments. 

We also use the number of distinct users for which the query was issued by, and use them as $o_{q}$ to calculate retrievability scores (see Section 4.3). Regarding the columns, as seen in Table~\ref{table:dataset_statistics}, the ones with numbers 1--3 are considered to be \narrowcols{}, whereas 4--9 are \broadcols{}. The broad columns which are free-text (\broadcolsft{}) are: \textit{Episode \& show description} and \textit{Transcript} for \podcasts{}, and \textit{User reviews} and \textit{Description} for \books{}.

Since the Goodreads dataset does not have any set of queries available, we generate a set of queries automatically: 75\% of the queries are narrow, generated by sampling words from the \narrowcols{}, and 25\% of them are from \broadcols{} and consider that as the relevance labels. This specific split of narrow and broad queries was chosen to simulate actual user behavior observed in the other two datasets (\tracks{} and \podcasts{}) where narrow queries are the majority but in a less extreme fashion. We use the number of ratings the entities from \books{} have as a proxy for the number of users that would issue such queries ($o_{q}$). 

\subsubsection{Broad queries datasets}

Since the majority of the queries from \tracks{}, \podcasts{} and \books{} are \narrow{}, we also employ two smaller additional sets of queries and relevance labels that have an underlying \broad{} intent. They are $\tracks{}_{\broad{}}$ and $\podcasts{}_{\broad{}}$, containing a total of 1309 and 500 queries. The $\tracks{}_{\broad{}}$ is a sample of queries from the logs that have a high predicted probability of being broad based on the interaction signals the user had after issuing the query. If the user interacts with entities such as playlists and hubs more than tracks and albums they are more likely to be issuing a broad query. Based on this set, we get the clicked entities where the query does not match the title, artist, or album of the entity, avoiding cases where the query seems broad but is in fact a narrow interaction, e.g. query ``\textit{pop}'' and clicking a track with the title ``\textit{POP!}''. For $\podcasts{}_{\broad{}}$, there is no parallel for the \tracks{} playlists so we employ a set of manually curated pairs of broad queries and entities. Annotators were instructed to write a query relevant to the podcast episode while avoiding exact matches and matching diverse metadata fields.

\subsection{Implementation Details}

\subsubsection{Query generation models}
As baselines for generating synthetic queries, we first use \textbf{\qgen{}}, a common approach to generate queries from documents used in this manner in different previous work~\cite{nogueira2019doc2query,liang2020embedding,ma2020zero,wang2021gpl}. We rely on fine-tuning T5~\cite{raffel2019exploring} (\textit{t5-base}) on a subset of \clicks{} train set with 10k pairs query-entities. The second baseline for generating queries requires very little supervision signal: \textbf{\inpars{}}~\cite{bonifacio2022inpars}. The model uses in-context learning, i.e. few examples in the prompt of the document and expected query, and large language models. For a fair comparison, we randomly sample examples to use in the prompt every time we are generating the output queries, this way \inpars{} has access to the same amount of training pairs of query and entities as \qgen{}\footnote{This was shown to be effective for the validation sets of \podcasts{} and \books{}. For \tracks{} we did not observe the same, so we used a fixed prompt with the same two examples randomly selected from the dataset.}. We rely on the open \textit{bigscience/bloom-760m}\footnote{\href{https://bigscience.huggingface.co/blog/bloom}{https://bigscience.huggingface.co/blog/bloom}} release to do so\footnote{We explore larger GPT-3 models on the appendix and see that the larger 175B parameter one does not significantly improve over smaller models.}. For the \textbf{\cqg{}} implementation we also rely on the T5 (\textit{t5-base}) model. When generating the queries with T5, for both \qgen{} and \cqg{} we employ $do\_sample$=True and $top\_k$=10.

\subsubsection{Retrieval models}
For \textbf{\bm{}}~\cite{robertson1994some} we resort to the default hyperparameters and implementation provided by the PyTerrier toolkit~\cite{pyterrier2020ictir}. For the zero-shot \textbf{\biencoder{}} models, we rely on the SentenceTransformers~\cite{reimers-2019-sentence-bert} model releases\footnote{\url{https://www.sbert.net/docs/pretrained_models.html}}. The library uses Hugginface transformers for the pre-trained models such as BERT~\cite{devlin2018bert} and MPNet~\cite{song2020mpnet}. Specifically, we employ the pre-trained model \textit{all-mpnet-base-v2}. When fine-tuning the \biencoder{} models on the \clicks{} or synthetic datasets, we rely on the \textit{MultipleNegativesRankingLoss}, which uses in-batch random negatives to train the model. We fine-tune the dense models for a total of 10k steps. \textbf{Thus, all dense models were trained on the same amount of (synthetic or not) queries}. We use a batch size of 8, with 10\% of the training steps as warmup steps. The learning rate is 2e-5 and the weight decay is 0.01. We refer to the \biencoder{} model trained on \clicks{} data as \textbf{$\biencoder{}_{\clicks{}}$} and a \biencoder{} model trained on the queries from \cqg{} as \textbf{$\biencoder{}_{\cqg{}}$}.

\subsection{Evaluation Procedure}

To evaluate the effectiveness of the retrieval systems we use the recall at 100, $R@100$. The choice for R@100 is due to the objective of increasing the retrievability of items considering the first 100 options\footnote{A second stage re-ranker in this pipeline could be precision-focused if the retriever is able to find enough relevant and diverse options.}. We perform Students t-tests at the confidence level of 0.95 with Bonferroni correction to compare the difference between models with statistical significance. 

To evaluate how biased the retrieval system is in terms of retrievability, we first estimate the retrievability of an entity $e$ as defined by~\cite{azzopardi2008retrievability}: $r(\mathbf{e})=\sum_{\mathbf{q} \in \mathbf{Q}} o_{q} \cdot f\left(k_{e q}, c\right)$, where $\mathbf{Q}$ is the set of queries\footnote{The size of $\mathbf{Q}$ is 100k for all computations.}, $o_{q}$ is the weight of each query---here we use the number of users that issued the query---and $f\left(k_{e q}, c\right)$ is 1 if the entity $e$ is ranked above $c$ by the search system (in our experiments we set c=100) and 0 otherwise. In order to get a number that summarizes how concentrated or biased the retrievability scores are we calculate the Gini score~\cite{gastwirth1972estimation}: $G=\frac{\sum_{i=1}^{N}(2 * i-N-1) * r\left(\mathbf{e}_{\mathbf{i}}\right)}{N \sum_{j=1}^{N} r\left(\mathbf{e}_{\mathbf{j}}\right)}$, where G=1 means only one entity concentrates all the retrievability, and G=0 means every entity in the collection has the same retrievability score. In order to perform statistical testing for the Gini scores we follow~\cite{gamboa2010statistical}.
\begin{table*}[ht!]
\caption{Retrieval effectiveness (R@100$\uparrow{}$ the higher the better) and retrievability bias (Gini $\downarrow{}$ the lower the better) of dense retrieval models trained on different training data for predominantly \narrow{} queries (\clicks{} test sets). Bold indicate the best model for each category with statistical significance and superscripts indicate statistically significant improvements over the respective model using students t-test at 0.95 confidence with Bonferoni correction for multiple comparisons. The values for the \books{} dataset on row ($c$) are not included as they are already a synthetic set of queries.
}
\label{table:results_narrow}
\begin{tabular}{@{}lp{1.5cm}p{1.5cm}p{1.5cm}p{1.5cm}p{1.5cm}p{1.5cm}@{}}
\toprule
 & \multicolumn{3}{c}{\textbf{R@100$\uparrow{}$}} & \multicolumn{3}{c}{\textbf{Gini$\downarrow{}$}} \\ \midrule
\begin{tabular}[c]{@{}l@{}}\textbf{Zero-shot}\\ \textit{(no target domain \clicks{} training data)}\end{tabular} & \tracks{} & \podcasts{} & \books{} & \tracks{} & \podcasts{} & \books{} \\ \midrule
(\textit{a}) \bm{} & 0.182$^{b}$ & 0.436$^{b}$ & \textbf{0.721}$^{bd}$ & 0.752$^{bh}$ & \textbf{0.666}$^{bcdefh}$ & \textbf{0.779}$^{b}$ \\
(\textit{b}) \biencoder{} & 0.142 & 0.323 & 0.415 & 0.818$^{h}$	& 0.765 & 0.836$^{d}$ \\
(\textit{c}) \biencoder{}$_{\weaklabelsun{}}$ (Ours) & \textbf{0.222} $^{abd}$ & \textbf{0.465}$^{b}$ & - & \textbf{0.748}$^{abh}$ & 0.730$^{bh}$ & - \\ \midrule
\multicolumn{7}{l}{\begin{tabular}[c]{@{}l@{}}\textbf{Fine-tuned on synthetic data} \\ \textit{(target domain \clicks{} training data to train query generators)}\end{tabular}} \\ \midrule
(\textit{d}) \biencoder{}$_{\inpars{}}$ \cite{bonifacio2022inpars} & 0.202$^{ab}$ & 0.474$^{ab}$ & 0.492$^{b}$ & 0.712$^{abch}$ & 0.677$^{bch}$ & 0.842 \\
(\textit{e}) \biencoder{}$_{\qgen{}}$ \cite{ma2020zero} & 0.296$^{abcd}$  & \textbf{0.503}$^{abc}$ & 0.755$^{abd}$ & 0.701$^{abcdh}$ & \textbf{0.674$^{bcdfh}$} & 0.766$^{abgh}$\\
(\textit{f}) \biencoder{}$_{\cqg{}}$ (Ours) & \textbf{0.333}$^{abcde}$ & 0.500$^{abc}$ & \textbf{0.770}$^{abde}$ & \textbf{0.693}$^{abcdeh}$ & 0.676$^{bdch}$ & \textbf{0.762}$^{abegh}$\\ \midrule
\multicolumn{7}{l}{\begin{tabular}[c]{@{}l@{}}\textbf{Fine-tuned on target data or in combination with synthetic data}\\ \textit{(access to target domain \clicks{} training data)}\end{tabular}} \\ \midrule
(\textit{g}) \biencoder{}$_{\clicks{} + \cqg{}}$ (Ours) & 0.361$^{abcdef}$ & 0.622$^{abcdef}$ & \textbf{0.775}$^{abde}$ &\textbf{0.817}$^{bh}$ & \textbf{0.741}$^{bh}$ & 0.768 $^{ab}$\\ 
(\textit{h}) \biencoder{}$_{\clicks{}}$ & \textbf{0.366}$^{abcdef}$ & \textbf{0.634}$^{abcdef}$ & 0.769$^{abde}$ & 0.856 & 0.763$^{b}$ & \textbf{0.767}$^{abg}$ \\ \bottomrule
\end{tabular}
\end{table*}

\section{Results}
In this section, we first describe the experimental results on \textbf{H1}---training dense retrieval models on synthetic queries leads to less retrievability bias than training on real queries and clicked entities---followed by the results for \textbf{H2}---suggesting broad queries generated by our proposed method \cqg{} leads to less retrievability bias when compared to the set of queries from the logs.

\subsection{H1: Modifying the Ranker with Generated Queries as Training Data}
\subsubsection*{\textbf{Evaluation with narrow intent queries}} 
Table~\ref{table:results_narrow} displays R@100 and Gini scores for different retrieval models on the three datasets which contain mostly narrow intent queries. Zero-shot models do not have access to any \clicks{} relevance labels for training. As expected, a \biencoder{} that has no access to the target domain queries and entities does not perform well, and it has worse effectiveness than \bm{} (30\% less R@100 on average, as seen row \textit{a} vs row \textit{b}). When using the target training data to fine-tune the dense retrieval model (\biencoder{}$_\clicks{}$) we observe that it outperforms zero-shot models significantly, with absolute gains of R@100 up to 158\% (row \textit{h} vs row \textit{b}). However, both the model trained on the \clicks{} data (row \textit{h}) and the pre-trained \biencoder{} (row \textit{c}) have significantly more bias than \bm{}, as seen by the Gini scores increases of 9.2\% and 10\% respectively.

When using the synthetic queries created by any of the query generation models to train the dense retrieval methods (\inpars{}, \qgen{}, \cqg{}) as described in Section 3.2.1, we observe significant drops of 10\% Gini on average (rows \textit{d},\textit{e},\textit{f} vs row \textit{h}), \emph{\textbf{indicating positive evidence for our first hypothesis}} that a model trained on the synthetic queries lead to less retrievability bias than the model trained on the \clicks{} data. Specifically, with \cqg{}\footnote{We employ here \cqg{}$_{narrow}$ which sets ($P_{narrow}$, $P_{broad}$) as ($100\%$, $0\%$) and \weaklabelsip{} as found to be optimal in the validation experiments (c.f. Table~\ref{table:ablation_study}).} we show that we can get statistically significant better effectiveness and retrievability for \tracks{} and \books{} than the query generation baselines with 24\% more R@100 and 3\% less Gini on average over all datasets and baselines (row \textit{f} vs rows \textit{d}, \textit{e}). We also show that we improve the retrievability over the model trained on \clicks{} data (row \textit{f} vs row \textit{h}) by 10\% Gini. This effectively makes more than 62k (9\%) entities in the \tracks{} dataset retrievable compared to $\biencoder{}_{\clicks{}}$, i.e. the entity goes from zero to a non-zero value.

We see also that with a combination of synthetic queries from \cqg{} and queries from the \clicks{} dataset~\footnote{We set the percentage as the optimal one in the validation set: 10\% synthetic queries and 90\% queries and clicks from \clicks{}.} (row \textit{g}) we can achieve similar effectiveness to the model training on the \clicks{} dataset (no statistical difference) while having less retrievability bias for both \tracks{} and \podcasts{} datasets with statistical significance, being Pareto optimal when considering both objectives.


\begin{table*}[]
\caption{Retrieval effectiveness (R@100$\uparrow{}$ the higher the better) and retrievability bias (Gini $\downarrow{}$ the lower the better) of dense models trained on different training data for a subset of \broad{} queries. Bold indicates the best model for each category with statistical significance and superscripts indicate statistically significant improvements over the respective model using Students t-test at 0.95 confidence with Bonferoni correction. When the set of queries $\set{Q'}$ were generated with ($P_{narrow}$, $P_{broad}$) as ($100\%$, $0\%$), we refer to it as $\cqg{}_{narrow}$, with ($0\%$, $100\%$) we call it $\cqg{}_{broad}$ and with ($50\%$, $50\%$) as $\cqg{}_{both}$.}
\label{table:gini_broad}
\begin{tabular}{@{}lllll@{}}
\toprule
\textbf{Method} & \multicolumn{2}{c}{\textbf{\textbf{R@100 $\uparrow$}}} & \multicolumn{2}{c}{\textbf{\textbf{Gini $\downarrow$}}} \\
& \tracksbroad{} & \podcastbroad{} & \tracksbroad{} & \podcastbroad{} \\\midrule
(\textit{a}) \biencoder{}$_{\cqg{}_{broad}}$  & \textbf{0.074}$^{bcdef}$ & 0.800$^{ef}$ & 0.596$^{b}$ & 0.831$^{bf}$\\
(\textit{b}) \biencoder{}$_{\clicks{}}$ & 0.035$^{def}$ & 0.756$^{f}$ & 0.878 & 0.846\\
(\textit{c}) \biencoder{}$_{\cqg{}_{both}}$ & 0.033$^{def}$ & 0.780$^{f}$ & 0.492 $^{abef}$ & 0.831$^{bf}$\\
(\textit{d}) \biencoder{}$_{\inpars{}}$  \cite{bonifacio2022inpars} & 0.010 & \textbf{0.827}$^{cef}$ & \textbf{0.489}$^{abcef}$ & \textbf{0.816}$^{abcf}$\\
(\textit{e}) \biencoder{}$_{\qgen{}}$ \cite{ma2020zero} & 0.009 & 0.744$^{f}$ & 0.540$^{ab}$ & 0.820$^{abcf}$\\
(\textit{f}) \biencoder{}$_{\cqg{}_{narrow}}$ & 0.003 & 0.609 & 0.517$^{abe}$ & 0.835$^{b}$ \\ \bottomrule
\end{tabular}
\end{table*}

\subsubsection*{\textbf{Evaluation with broad intent queries}} 
In order to understand how the models perform for exploratory and complex information needs, we take a closer look at the effectiveness and retrievability of the models in a set containing only broad intent queries. Table~\ref{table:gini_broad} shows that a model trained on synthetic queries from \cqg{} gets significantly better when we include broader queries in the training (going from 0, 50 and 100\% on rows \textit{f}, \textit{c} and \textit{a}). 

A model trained only on synthetic broad queries outperforms a model trained on \clicks{} data by 111\% of R@100 for \tracksbroad{} with statistical significance (row \textit{a} vs row \textit{b}). We also observe significant drops in the retrievability bias when we compare models trained with \cqg{} queries with models trained with \clicks{} data, going from 0.878 to 0.596 and from 0.846 to 0.831 as seen in Table~\ref{table:gini_broad} (row \textit{a} vs row \textit{b}). Baseline models to generate queries (rows \textit{d} and \textit{e}) also have significantly less retrievability bias than the model trained on click data (row \textit{b}) \textbf{\emph{showing again positive evidence for our first hypothesis}} that models trained on synthetic queries lead to less retrievability bias.

\begin{table*}[]
\centering
\caption{Ablation study where we add components of the proposed \cqg{} to \qgen{} one at a time using \clicks{} validation sets which has predominantly narrow queries. We also report the effectiveness for the \broad{} queries only datasets. The $\dagger$ superscripts indicate statistically significant improvements over \qgen{} using students t-test at 0.95 confidence.}
\label{table:ablation_study}
\begin{tabular}{@{}llllllllllll@{}}
\toprule
& & \multicolumn{6}{c}{Predominantly narrow queries}& \multicolumn{4}{c}{Broad queries}  \\ \midrule
& & \multicolumn{3}{c}{\textbf{R@100}$\uparrow{}$}& \multicolumn{3}{c}{\textbf{Gini} $\downarrow$}& \multicolumn{2}{c}{\textbf{R@100}$\uparrow{}$} & \multicolumn{2}{c}{\textbf{Gini} $\downarrow$} \\ \cmidrule(l){3-12} 
& & \multicolumn{1}{c}{\tracks{}} & \multicolumn{1}{c}{\podcasts{}} & \multicolumn{1}{c}{\books{}} & \multicolumn{1}{c}{\tracks{}} & \multicolumn{1}{c}{\podcasts{}} & \multicolumn{1}{c}{\books{}} & \multicolumn{2}{c}{\tracksbroad{}} & \multicolumn{2}{c}{\podcastbroad{}}\\ \midrule
\multicolumn{2}{l}{\qgen{} \cite{ma2020zero}}  & 0.289 & 0.512 & 0.756 & 0.701 & 0.674 & 0.766 & 0.009 & 0.744 & 0.540 & 0.820\\ \hdashline
\multicolumn{2}{r}{} &  &  &  & &  &   &  & & & \\
\multicolumn{2}{l}{+ (1) Serialization}   & \textbf{0.312}$^{\dagger}$ & 0.509 & \textbf{0.761}$^{\dagger}$ & \textbf{0.694}$^{\dagger}$ & \textbf{0.666}$^{\dagger}$ & \textbf{0.761}$^{\dagger}$ & 0.006 & 0.711 & \textbf{0.528}$^{\dagger}$ & 0.824 \\
\multicolumn{2}{r}{} &  &  &  & &  &   &  & & & \\
\multicolumn{2}{l}{+ (2a,3) Intent-aware Generation} & \textbf{0.305}$^{\dagger}$ & 0.505 & \textbf{0.761}$^{\dagger}$ & \textbf{0.688}$^{\dagger}$ & \textbf{0.669}$^{\dagger}$ & \textbf{0.756}$^{\dagger}$  & 0.008 & 0.751 & \textbf{0.528}$^{\dagger}$ & 0.820\\
\multicolumn{2}{r}{ $^{\weaklabelsip{}}$ } &  &  &  & &  &   &  & & &\\
\multicolumn{2}{l}{+ (2a,2b,3) Intent-aware Generation} & 0.289 & 0.508 & - & 0.704 & 0.704 & -  & \textbf{0.036}$^{\dagger}$ & \textbf{0.787}$^{\dagger}$ & \textbf{0.486}$^{\dagger}$ & 0.829\\
\multicolumn{2}{r}{ $^{\weaklabelsip{} + \weaklabelsun{}}$} &  &  &  & &  &   &  & & &\\
\multicolumn{2}{l}{+ (1,2a,3) \cqg{}} & \textbf{0.300}$^{\dagger}$ & \textbf{0.522}$^{\dagger}$ & \textbf{0.763}$^{\dagger}$ & \textbf{0.694}$^{\dagger}$ & 0.676 & \textbf{0.760}$^{\dagger}$ & 0.007 & 0.713 & \textbf{0.522}$^{\dagger}$ & 0.822 \\ \multicolumn{2}{r}{ $^{\weaklabelsip{}}$ } &  &  &  & &  &   &  & & &\\
\multicolumn{2}{l}{+ (1,2a,2b,3) \cqg{}} & 0.283 & 0.490 & - & 0.701 & 0.674 & -  & \textbf{0.033}$^{\dagger}$ & \textbf{0.780}$^{\dagger}$ &  \textbf{0.480}$^{\dagger}$ & 0.833 \\
\multicolumn{2}{r}{ $^{\weaklabelsip{} + \weaklabelsun{}}$} &  &  &  & &  &   &  & & &\\
\bottomrule
\end{tabular}
\end{table*}
\subsubsection*{\textbf{Contribution of each module of \cqg{}}}
When using the set of queries generated by \weaklabelsun{}(no supervision available) to train the dense retriever, we get statistically significant improvements over the model that is not fine-tuned (row \textit{c} vs row \textit{b}, going from 0.142 to 0.222 R@100 and from 0.323 to 0.465 R@100 for the \tracks{} and \podcasts{} datasets as seen in Table~\ref{table:results_narrow}. Using the remaining components of \cqg{} we obtain significant improvements over using solely \weaklabelsun{}. A natural question is from which modules the improvements are coming. To answer this question Table~\ref{table:ablation_study} displays an ablation study on the components of \cqg{}. If we remove all components of \cqg{} we end up with the baseline \qgen{} (first line of the table). We incrementally add each component of the model in the following rows. 

The main findings are that (I) the serialization component, where we indicate which the metadata columns and their respective values as opposed to values only, is beneficial for both R@100 and Gini for both the narrow evaluation set of queries and broad set of queries and (II) the \weaklabelsun{} is only beneficial for the broad set of queries as \weaklabelsip{} cover narrow queries well (they are the majority of the available existing queries). 


\subsection{H2: Modifying the Set of Queries by Suggesting Generated Queries}


In order to test our second hypothesis that suggesting broad queries with \cqg{} leads to less retrievability bias compared to the queries found in the logs, we rely on a simulation where a percentage of suggested queries are accepted and added into the set of queries as described in Section 3.2.2. For each entity in the top-5 ranked list for the log queries we create 3 query suggestions.

\begin{figure}[]
    \centering
    \includegraphics[width=.45\textwidth]{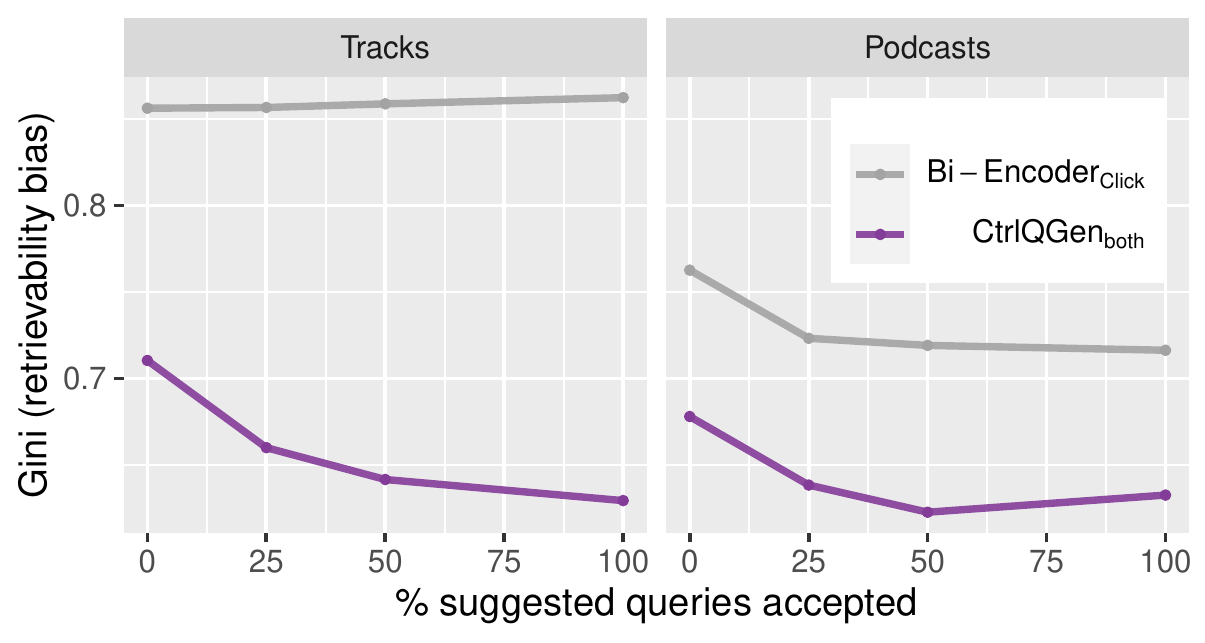}
    \caption{Suggesting broad queries with \cqg{} reduces the retrievability of a model trained on synthetic data.}
    \label{fig:query_suggestion}
    \vspace{-0.5cm} 
\end{figure}

Figure~\ref{fig:query_suggestion} displays the results of this simulation. We see that if the \biencoder{} is trained on a set of broad queries, the retrievability of the system drops significantly as higher percentages of suggested broad queries by \cqg{} are accepted, with decreases of Gini up to 11\% and 7\% for \tracks{} and \podcasts{} \textbf{\emph{showing positive evidence for our second hypothesis}}. If we consider that all queries are accepted by the users a total of 78k (11\%) entities for the \tracks{} dataset would become retrievable, i.e. retrievability different than zero, compared to using \cqg{}$_{both}$ with the log queries. We also see that only modifying the set of queries is not enough, as a Bi-Encoder trained on the \clicks{} data does not achieve the same effect, showing that it is also necessary to employ a model that was trained for both narrow and broad queries.


\section{Conclusion}

We propose here \cqg{}, a new approach to generate synthetic queries for entities that allows to control for the query intent and that can work in the absence of annotated data through the use of a weak-labeling function that leverages content metadata. We study the impact that the generated queries have on decreasing the retrievability bias and effectiveness, i.e. on helping the search engine surface more entities while avoiding negative effects on the relevance of results. 
Our experimental results in three different domains show that training dense retrieval models on synthetic queries from \cqg{} leads to significant decreases in the retrievability bias of the system with comparable effectiveness. We also demonstrate how to reduce the retrievability bias by suggesting queries generated by \cqg{}.

As future work, we believe important directions to be: (I) taking into account the interplay between recommendation and search in the measure of the accessibility of an entity, (II) improving the representation of entities for which most metadata information is not available and (III) study methods to reduce the retrievability of a system for re-ranking scenarios (IV) study the impact of increased content retrievability on content discovery.


\bibliographystyle{ACM-Reference-Format}
\bibliography{sample-base}

\appendix
\section{Unsupervised Weak Labeling Functions}
In this appendix, we define the functions used in \weaklabelsun{}.
\subsection{Random Terms Selection}
Samples words from the entity $e$, given possible $P$ length percentages for the query with probability $Pr$.
\begin{python}
def sample_words(e, P, Pr):
    words = tokenize(e)
    p_words_to_sample = np.random.choice(P, 1, p=Pr)
    n = int(len(words) * p_words_to_sample)
    words = np.random.choice(words, n, replace=False)
    return words
\end{python}

\subsection{Query Variation Ordering}
Generates a query variation by shuffling two random words from the query.
\begin{python}
def qv_ordering(q):
    words = tokenize(q)
    idxs = [i for i in range(0, len(words))]
    p1, p2 = np.random.choice(idxs, 2, replace=False)
    words[p1], words[p2] = words[p2], words[p1]
    return " ".join(words)
\end{python}

\subsection{Query Variation Misspelling}
Generates a query variation by adding a misspelling error with $P$ probabilities of removing and addition.
\begin{python}
def qv_misspelling(q, P):
    t = np.random.choice(["rem", "mdf"], 1, p=P)
    idxs=[i for i in range(len(query))]
    l=string.ascii_letters
    if t == "rem":
        idx_rem = np.random.choice(idxs, 1)[0]
        qv = q[0:idx_rem] + q[idx_rem+1:]
    elif t == "mdf":
        idx_mdf = np.random.choice(idxs, 1)[0]
        char_add = np.random.choice(len(l), 1)[0]
        qv = q[0:idx_mdf] + l[char_add] + q[idx_mdf+1:]
    return qv
\end{python}

\subsection{Query Variation Prefix}
Generates a query variation by removing $P$ percentages of the suffix of the query with probabilities $Pr$.
\begin{python}
def qv_prefix_query(q, P, Pr):
    rem = np.random.choice(P, 1, p=Pr)    
    return q[:int((1-rem)*len(q))]
\end{python}

\subsection{Query from Free-Text Column by Summarization}
Generates a query by summarizing the value of a free-text column (\broadcolsft{}). For our experiments we rely on the pre-trianed summarizer model \textit{snrspeaks/t5-one-line-summary}\footnote{\url{https://huggingface.co/snrspeaks/t5-one-line-summary}}.
\begin{python}
from transformers import pipeline
def q_summarizer(ft, m):
    pipe = pipeline("text2text-generation", 
                    model = m)
    return pipe("summarize: {}".format(ft))[0]
\end{python}



\section{Bias mitigation for the \clicks{} dataset}
In this appendix, we investigate if it is possible to mitigate the biases of the \clicks{} data with a simpler approach.

When fine-tuning the Bi-Encoder with \clicks{} data in our experiments we do not employ the same combination of queries and entities twice, even if that pair is highly popular in the logs. This is already a way of reducing the bias in the \clicks{} dataset. However, there are still many query variations that lead to the same entities, i.e. queries with different forms but with the same underlying information need, which are not removed when we get distinct queries for training. In order to mitigate this bias from the \clicks{} data by removing multiple queries that lead to the same entity, we randomly select only one of the queries for each entity to train the \biencoder{} model on. 

The result of this experiment is that such a bias mitigation strategy indeed improves the retrievability of the system: the Gini scores go from 0.856 to 0.803 for \tracks{} and from 0.763 to 0.713 for \podcasts{}. However the mitigated \clicks{} data approach still leads to 30\% and 5\% more retrievability bias than \cqg{}, for \tracks{} and \podcasts{} respectively.

\section{Scaling \inpars{} with GPT-3}
In this appendix, we test if increasing the model size of the InPars model using GPT-3 as the language model has a significant effect on the Bi-Encoder trained with such synthetic queries. 

\begin{figure}[H]
    \centering
    \includegraphics[width=.45\textwidth]{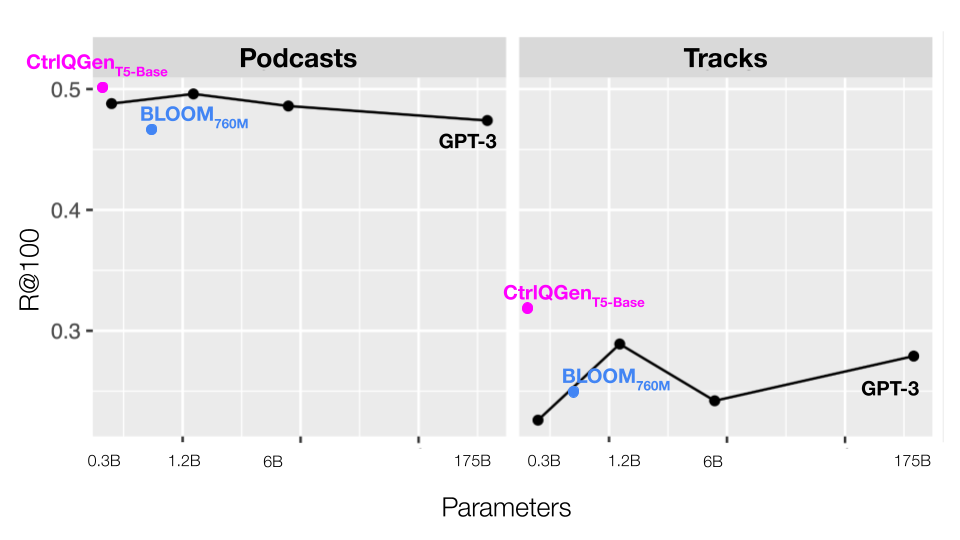}
    \caption{Scaling \inpars{} baseline using GPT-3.}
    \label{fig:gpt3}
\end{figure}

We see from \ref{fig:gpt3} that this is not the case for both datasets, and the highest R@100 is reached when using the 1.2B GPT-3 model (\textit{babbage-001}). Similar results were found in the InPars paper~\cite{bonifacio2022inpars}.

\section{Overlap of generated queries}
In this appendix, we check if the queries generated by \cqg{} have a significant overlap with either the log queries (is \cqg{} just copying existing queries?) or with the input entity (is \cqg{} just copying words from the input entity?).

\subsection{With log queries from \clicks{}}
Out of the 10k narrow queries generated by 
\cqg{} to test the first hypothesis (results from Tables~\ref{table:results_narrow}), there are only 6\% and 12\% exact matches with set of queries from the logs (\clicks{}) for the \tracks{} and \podcasts{} datasets respectively. For the second hypothesis, out of the 376k broad queries generated, there are only 2\% and 1\% are exact matches with the set of queries from the logs. This shows the diversity of the generated queries from \cqg{}, and that it is not just copying input queries from the log.

\subsection{With the input entity}
Out of the 10k narrow queries generated by 
\cqg{} to test the first hypothesis, 25\% and 48\% are not a subset of the serialized entity for \tracks{} and \podcasts{} datasets respectively. For the second hypothesis, out of the 376k broad queries generated, a total of ~70\% of queries are not subsets of the serialized entity for both datasets. This shows that while for narrow queries substrings of the entity are the majority of the cases when generating broad queries this is not the case. Also, this indicates that for both cases the model is not always selecting parts of the input as the query.

\section{Dataset details}

For the \books{} dataset, we take into account the top two most-voted reviews and use the first 50 tokens. For the \tracks{} dataset, we use the first lyric line and the most frequent lyric line and employ a maximum of 25 descriptors. For the \podcasts{} dataset, we use the first 50 tokens of the description and of the transcript. For the \books{} and \tracks{} datasets we use a maximum of 25 playlists.

\end{document}